\documentclass[final]{svjour2}
\usepackage{graphicx}
\usepackage{rotating}
\usepackage{amssymb}
\usepackage{mathptmx}
\usepackage[numbers]{natbib}

\newcommand{\barray}{\begin{eqnarray}}
\newcommand{\earray}{\end{eqnarray}}

\newcommand{\beq}{\begin{equation}}
\newcommand{\eeq}{\end{equation}}
\newcommand{\ba}{\begin{array}}
\newcommand{\ea}{\end{array}}
\newcommand{\bea}{\begin{eqnarray}}
\newcommand{\eea}{\end{eqnarray} }
\newcommand{\be}{\begin{eqnarray}}
\newcommand{\ee}{\end{eqnarray} }
\newcommand{\bal}{\begin{align}}
\newcommand{\eal}{\end{align}}
\newcommand{\bi}{\begin{itemize}}
\newcommand{\ei}{\end{itemize}}
\newcommand{\ben}{\begin{enumerate}}
\newcommand{\een}{\end{enumerate}}
\newcommand{\bc}{\begin{center}}
\newcommand{\ec}{\end{center}}
\newcommand{\bt}{\begin{table}}
\newcommand{\et}{\end{table}}
\newcommand{\btb}{\begin{tabular}}
\newcommand{\etb}{\end{tabular}}

\makeatletter
\journalname{Journal of Low Temperature Physics}

\bibpunct{}{}{,}{s}{}{,}

\begin{document}

\newcommand{\hdblarrow}{H\makebox[0.9ex][l]{$\downdownarrows$}-}
\title{Higgs bosons in particle physics and in condensed matter}

\author{G.E. Volovik$^{1,2}$ \and M.A. Zubkov$^3$}

\institute{1: O.V. Lounasmaa Laboratory, School of Science and Technology,
Aalto University, Finland\\
Tel.: +358 50 344 2858\\ Fax: +358 9 470 22969\\
\email{volovik@ltl.tkk.fi}
\\2: L.D. Landau Institute for Theoretical Physics, Moscow, Russia\\ 3:
Institute for Theoretical and Experimental Physics, Moscow, Russia}

\date{30.05.2013}

\maketitle

\keywords{Higgs boson, Standard Model, Goldstone modes, superfluid $^3$He}

\begin{abstract}

Higgs bosons -- the amplitude modes -- have been experimentally investigated
in condensed matter for many years.
An example is superfluid $^3$He-B, where the broken symmetry leads to 4
Goldstone modes and at least 14 Higgs modes, which are characterized by
angular momentum quantum number $J$ and parity (Zeeman splitting of Higgs
modes with $J=2^+$ and $J=2^-$  in magnetic field has been  observed in
80's).  Based on the  relation $E_{J+}^2+E_{J-}^2=4\Delta^2$ for the energy
spectrum of these modes, Yoichiro Nambu proposed the general sum rule, which
relates masses of Higgs bosons and masses of fermions. If this rule is
applicable to Standard Model, one may expect that  the observed Higgs boson
with mass $M_{{\rm H}1}=125$ GeV  has a Nambu partner -- the second Higgs
boson with mass $M_{{\rm H}2}=325$ GeV. Together they satisfy the Nambu
relation $M_{{\rm H}1}^2 + M_{{\rm H}2}^2 = 4 M_{\rm top}^2$, where $M_{\rm
top}$  is the top quark mass.
Also the properties of the Higgs modes in superfluid $^3$He-A, where
the symmetry breaking is similar to that of the Standard Model, suggest the
possible existence of two electrically charged Higgs particles
with masses  $M_{{\rm H}+}=M_{{\rm H}-}\sim 245$ GeV, which together obey
the Nambu rule
$M_{{\rm H}+}^2 + M_{{\rm H}-}^2 = 4 M_{\rm top}^2$. A certain excess of
events at 325 GeV and at 245 GeV has been reported in 2011, though not
confirmed in 2012 experiments. Besides, we consider the particular
relativistic model of top - quark condensation that suggests the possibility
that two twice degenerated Higgs bosons contribute to the Nambu sum rule.
This gives the mass around $210$ GeV for the Nambu partner of the $125$ GeV
Higgs boson.
We also discuss the other possible lessons from the condensed matter to
Standard Model, such as hidden symmetry, where light Higgs emerges as quasi
Nambu-Goldstone mode, and the role of broken time reversal symmetry.

PACS numbers: 14.80.Ec, 67.30.-n
\end{abstract}

\section{Introduction}

Condensed matter physics and  particle physics use the same methods of
quantum field theory and operate with similar phenomena. Typical example is
the Anderson-Higgs mechanism of the formation of mass of gauge bosons, which
has been discussed both in Standard Model of particle physics (SM) and in
superconductors, where gauge symmetry is spontaneously broken. The gauge
bosons become massive due to entanglement with Nambu-Goldstone (NG) bosons
\cite{Anderson1963,Englert,Higgs,Kibble}.

The Higgs amplitude modes -- known as Higgs bosons --  represent the other
common objects. They have been first discovered in condensed matter:  in
superfluid $^3$He \cite{Lawson1973,Paulson1973} and later in superconductors
\cite{LittlewoodVarma1981}. The discovery of the first Higgs boson in
particle physics generated the new interest to their counterparts in
condensed matter, see e.g. recent papers
\cite{Endres2012,Gazit2012,Barlas2013,KunChen2013,Matsunaga2013} and
references therein.
We concentrate here mainly on Higgs bosons in superfluid $^3$He, which were
studied for many years
theoretically and experimentally and were served as inspiration for particle
physics.

It was observed by Nambu \cite{Nambu1985}, that in systems described by the
BCS theory (superconductors,  nuclear matter and especially superfluid
$^3$He-B)
there is a remarkable relation between the masses of the fermions and
the masses of bosons. The collective bosonic modes emerging in the fermionic
system -- NG modes and
Higgs amplitude modes -- can be distributed into the pairs of Nambu
partners. For each
pair one has the relation,
\begin{equation}
 M_{1}^2 + M_{2}^2 = 4 M^2_f   \,,
\label{NambuRelation}
\end{equation}
where $M_1$ and $M_2$ are gaps in the
bosonic spectrum, and
 $M_f$ is the gap in the femionic spectrum. In relativistic systems gaps in
the energy spectrum corresponds
 to the mass of particles, which suggests that the masses of fermions and
Higgs bosons in the relativistic theories, such as SM, can be related.
Such relation exists for example in the Nambu - Jona - Lasinio (NJL) model
\cite{NJL} of quantum chromo-dynamics, where  it relates masses of  the
$\sigma$ - meson
and of the constituent quark $M_{\sigma} \approx 2 M_{quark}$.

We discuss the Nambu sum rule in $^3$He-B and in thin films of $^3$He-A in
Sec. \ref{Nambu}, with application to SM Higgs bosons. In Sec.
\ref{HiddenSymmetry} we discuss the effect of hidden symmetry, which leads
to the relatively small mass of  Higgs boson, which  emerges as a quasi NG
mode.
The role of flat directions in the Higgs potential is discussed in Sec.
\ref{FlatDirections}.
In Sec. \ref{GaplessFermions} the Nambu sum rule is extended to 3D $^3$He-A,
where the spectrum of fermions is anisotropic and gapless.
The role of broken time reversal symmetry in transformation of NG boson to
the Higgs boson
is discussed in Sec. \ref{TimeReversalSymmetry} on example of spin and
orbital waves in ferromagnets
and Kelvin waves on quantized vortices. Sec. \ref{NJL} is devoted to the
consideration of the relativistic NJL model of top - quark condensation,
where the Nambu sum rule naturally arises.

\section{Nambu sum rule for SM Higgs bosons: hints from superfluid $^3$He}
\label{Nambu}

\subsection{Higgs field and Higgs potential in superfluid $^3$He}

Superfluidity in liquid $^3$He and superconductivity are based on the
mechanism of Cooper pairing.
The  Higgs field  appears as a composite
object made of two fermions --  two $^3$He atoms in superfluid $^3$He or two
electrons in superconductors. The order parameter is the vacuum expectation
value of the creation operator of two fermions, such as   $\left<ee\right>$
for Cooper pairing of electrons in superconductors.

In superfluid $^3$He the condensate is formed by Cooper pairs  in the
spin-triplet
$p$-wave state. The order parameter (Higgs field) is $3\times 3$ complex
matrix $A_{\alpha i}$,
 it  transforms as a vector under a spin
rotation for given orbital index ($i\, $)
-- and as a vector under an orbital
rotation for given spin index ($\alpha $).
The Ginzburg-Landau free energy functional  -- the Higgs potential --  is
invariant under the group $G=SO_S(3) \otimes SO_L(3) \otimes U(1)$ of spin,
orbital and gauge rotations  \cite{VollhardtWolfle1990}:
\begin{eqnarray}
F = -\alpha A^{\ast }_{\alpha i} A_{\alpha i}
+\beta _1A^{\ast } _{\alpha i} A^{\ast } _{\alpha i}
A_{\beta j} A_{\beta j}
+
\beta _2A^{\ast }_{\alpha i}
A_{\alpha i} A^{\ast } _{\beta j} A_{\beta j}
\nonumber
\\
+\beta _3 A^{\ast } _{\alpha i} A^{\ast } _{\beta i}
A_{\alpha j} A_{\beta j}+ \beta _4 A^{\ast }_{\alpha i} A_{\beta i}
A^{\ast }_{\beta j}
A_{\alpha j}
+\beta _5 A^{\ast }_{\alpha i} A_{\beta i} A_{\beta j} A^{\ast }_{\alpha
j}\,.
\label{HiggsPotential}
\end{eqnarray}
The approximate symmetry $SO_S(3) \otimes SO_L(3)$  with respect to separate
spin and orbital  rotations  is similar to the so-called custodial symmetry
in particle physics.  It gives extra NG bosons in $^3$He-A and in $^3$He-B,
which become Higgs bosons with a relatively small mass (Leggett frequency)
due to
a tiny spin-orbit interaction.

\subsection{Higgs bosons in $^3$He-B}

The B-phase of $^3$He is characterized by the quantum numbers $S=1$, $L=1$,
$J=0$ of spin, orbital momentum and
total angular momentum respectively
 \cite{VollhardtWolfle1990}. This corresponds to the symmetry breaking
scheme $G \rightarrow  H$, where the symmetry of the degenerate vacuum
states is $H=SO_J(3)$.
The collective modes  in the vicinity of an equilibrium degenerate state,
chosen as $A_{\alpha i}({\rm eq})= \Delta \delta_{\alpha i}$ with $\Delta$
being the gap in the fermionic spectrum, are propagating  deviations of the
Higgs field
\begin{equation}
A_{\alpha i}- A_{\alpha i}({\rm eq}) = u_{\alpha i} +iv_{\alpha i} \,.
\label{B-phase}
\end{equation}
Altogether there are 18 real variables $u$ and $v$, and correspondingly 18
collective
bosonic modes. These modes classified by quantum numbers $J = 0,1,2$
have been calculated in the earlier papers
(see e.g. Refs. \cite{Vdovin1963,Maki1974,Nagai1975,Tewordt-Einzel1976}).
Four modes are gapless NG bosons resulting from
the symmetry breaking $G \rightarrow  H$. This satisfies the conventional
wisdom that the total number of NG modes
$=$ the number of broken symmetry generators ($7-3=4$).
The rest  14 bosons  are amplitude modes -- Higgs bosons with non-zero gaps.
The energy gaps of real and imaginary modes are related
 \cite{VolovikZubkov2013}:
\begin{equation}
E_{u,v}^{(J)} = \sqrt{ 2 \Delta^2(1\pm  \eta^{(J)})}  \,,
\label{SymmetryConsideration}
\end{equation}
where parameters $ \eta^{(J)}$ are determined by the symmetry of the system,
 $\eta^{J=0} = \eta^{J=1} = 1$,
and $\eta^{J=2} = \frac{1}{5}$. Thus the symmetry consideration supports the
Nambu
conjecture (\ref{NambuRelation}) for $^3$He-B: the gaps of Nambu partners in
each sector $J$ satisfy the Nambu rule
\begin{equation}
[E_{u}^{(J)}]^2 + [E_{v}^{(J)}]^2  = 4 \Delta^2 \,.
\label{NSRThe}
\end{equation}
The sector $J=0$ contains one pair of the Nambu partners (the Higgs
amplitude mode with gap $2\Delta$ -- the pair-breaking mode, and the NG
mode -- sound wave):
\begin{equation}
E^{(0)}_u = 2 \Delta,\quad E^{(0)}_v =  0.
\end{equation}
For $J=1$ there are 3  pairs (3 NG
modes --  spin waves, and 3 Higgs modes):
\begin{equation}
E^{(1)}_u = 0,\quad E^{(1)}_v =  2 \Delta.
\end{equation}
The sector  $J=2$ contains 10 Higgs bosons which form 5 Nambu pairs (5 real
squashing modes + 5
imaginary squashing modes):
  \begin{equation}
E^{(2)}_u = \sqrt{2/5}\, (2\Delta),\quad E^{(2)}_v =  \sqrt{3/5}\,
(2\Delta)\,.
\label{He3B2}
\end{equation}
The 5-fold Zeeman splitting of the Higgs modes with $J=2$  in magnetic field
has been  observed in 80's  \cite{Avenel1980,Lee1988}, for the latest
experiments  see  \cite{Collett2012}.

Equation (\ref{He3B2}) relating two Higgs bosons in $^3$He-B
may serve as a hint for SM. If the symmetry breaking in SM is related to
the top quark condensate $\left<\bar t t\right>$, then one may expect that
the discovered Higgs boson with mass $M_{{\rm H}1}=125$ GeV has the Nambu
partner with mass   $M_{{\rm H}2}=\sqrt{4M_{\rm top}^2-M_{{\rm H}1}^2}\sim
325$ GeV.
In 2011 the
CDF collaboration \cite{CDF} has announced the preliminary results on the
excess of events in $Z Z \rightarrow l l \bar{l} \bar{l}$ channel at the
invariant mass $\approx 325$ GeV.  CMS collaboration also reported a small
excess in this region \cite{CMS}.
In \cite{325,320} it was argued that this may point out
to the possible existence of  a new scalar particle with mass $M_{{\rm
H}2}\approx
325$ GeV.

\subsection{Higgs bosons in superfluid  phases in 2+1 films}

The Higgs field $A_{\alpha i}$ in 2D thin films contains $3\times 2\times
2=12$ real components.
There are two possible phases:  the A-phase and the planar phase.
Both phases have isotropic gap $\Delta$ in the 2D case. The degenerate
vacuum state
of the A-phase,
$A_{\alpha i}({\rm eq})= \Delta \hat z_\alpha(\hat x_i + i\hat y_i)$,
corresponds to the symmetry
breaking
$G=SO_L(2) \otimes SO_S(3) \otimes U(1) \rightarrow H=U(1)_Q\otimes
SO_S(2)$, where the combined symmetry $U_Q(1)$ is similar to the
electromagnetic symmetry of SM.   The 12 collective modes are classified in
terms of the ``electric'' charge $Q$ and include $5-2=3$  NG bosons + 9
Higgs
amplitude modes. Their energies obey Eq.(\ref{SymmetryConsideration}),
with quantum number $Q$ instead  of $J$. This is another example, where the
Nambu sum rule works.
The parameters $\eta$ are determined by the symmetry of the system. Both in
the A-phase and in the planar phase they get three possible values $\eta=1$,
$\eta= -1$, and $\eta = 0$.
In the A-phase,  these modes  form two pairs of Nambu partners (triply
degenerated), with $Q=0$ and $|Q|=2$ (see also  Ref.
\cite{BrusovPopov1981}):
\begin{eqnarray}
E_1^{(Q=0)} =0~~,~~ E_2^{(Q=0)} = 2 \Delta\,,
\label{2DA1}
\\
E^{(Q=+2)}=
\sqrt{2}\Delta~~,~ E^{(Q=-2)}= \sqrt{2}\Delta  \,.
\label{2DA2}
\end{eqnarray}
Since masses of $Q=+2$ and $Q=-2$ modes are equal,
the Nambu rule necessarily leads
to the definite value of the masses of the ``charged'' Higgs bosons.
Because of the common symmetry breaking scheme in SM and
in $^3$He-A, Eq.(\ref{2DA2}) may serve as a hint for
existence of two Higgs bosons in SM with equal masses
\begin{equation}
M_{{\rm H}+}=M_{{\rm H}-}=\sqrt{2} M_{\rm top}  \,.
\label{2D}
\end{equation}
This mass is about 245 GeV.
A certain excess of events in this region has been
observed by ATLAS
in 2011 (see, for example, \cite{ATLASHiggs}).

\section{Hidden symmetry: light Higgs as quasi Nambu-Goldstone mode}
\label{HiddenSymmetry}

The mass of the observed Higgs boson is rather small compared to the
characteristic
electroweak scale of order 1 TeV. This may indicate an existence of some
approximate (custodial symmetry). We have already mentioned the  custodial
symmetry of separate spin and orbital rotations  in superfluid $^3$He, which
leads
to Higgs bosons with small mass originating from the quasi-NG modes -- spin
waves.

Here we consider the hidden symmetry emerging in the BCS theory of
superfluid $^3$He-A, which corresponds to the weak coupling approximation.
Application of this hidden symmetry to the structure of the topological
defects  in $^3$He-A was discussed in \cite{MerminMineevVolovik}.
In the  BCS
approximation, there are the following relations between the
$\beta$-parameters of quartic terms in Higgs potential
(\ref{HiggsPotential}):
$-2\beta _1 = \beta _2 = \beta _3 =\beta _4 = -\beta _5$.
These relations have a crucial
effect for bosons in $^3$He-A: they give rise to 3 extra NG bosons   due
to hidden symmetry and one more NG boson due to flat direction
\cite{VolovikHazan,Exotic}.

The hidden symmetry can be visualized in the following way. The A-phase
Higgs field
$A_{\alpha i}({\rm eq})= \Delta \hat x_\alpha(\hat x_i + i\hat y_i)$ can be
represented as a sum of two terms
\begin{equation}
A_{\alpha i}({\rm eq})=
\frac{\Delta }{2}(\hat x_\alpha + i \hat y_\alpha) (\hat x_i + i\hat y_i)  +
\frac{\Delta }{2}   (\hat x_\alpha - i \hat y_\alpha) (\hat x_i + i\hat
y_i)\,.
\label{spin-up-spin-down}
\end{equation}
The first term represents the subsystem with quantum numbers $S_z=L_z=+1$
(spin-up component), while the second subsystem has $S_z=-L_z=-1$
(spin-down component).  In the BCS theory of $^3$He-A, the spin-up and
spin-down components of
Higgs field are independent: they may have different phases and different
directions of orbital quantization axis,  $\hat{\bf l}_+$ and $\hat{\bf
l}_-$.  Together with 2 degrees of freedom for the choice of  spin
quantization axis,
 the vacuum states of the Higgs field have
$(2+1)\times 2+2=8$ degrees of freedom. According to conventional wisdom,
this suggests 8  NG bosons instead of 5 NG modes in the absence
of custodial symmetry. Thus the hidden symmetry should lead to $8-5=3$
extra NG bosons, which acquire small mass due to quantum corrections and
become the Higgs fields.

This rule of counting of the number of NG bosons is obeyed for all $^3$He-A
vacua with one exception: on the sub-manifold of the vacuum states where the
orbital vectors $\hat{\bf l}_+$ and $\hat{\bf l}_-$ of the two spin
subsystems are  equal as in Eq.(\ref{spin-up-spin-down}), the number of NG
modes is 9 instead of 8, thus violating the conventional wisdom.

The theorems concerning  the number of of NG modes in the broken symmetry
states are  discussed in recent literature (see Refs.
\cite{WatanabeMurayama2012,WatanabeBraunerMurayama2013,Kapustin2012,Uchino20
10}
and references therein). With some nondegeneracy assumption about the
low-energy effective action,
the total number of NG bosons (or quasi-NG bosons, if the symmetry is
hidden)
adds up to the number of broken symmetry generators.
 Typically this is the difference between the number of
generators of $G$ and $H$ groups. The number of NG modes can be smaller,
e.g. if the time-reversal symmetry is
violated, see Sec. \ref{TimeReversalSymmetry}.

However, $^3$He-A provides an example where the number of NG modes exceeds
the  number of broken symmetry generators. Due to this example, the counting
rule has been reformulated  by S.P. Novikov:  the number of NG modes
coincides with the dimension
of the ``tangent space'' \cite{Novikov1982}. The mismatch between
the total number of NG bosons and the number
of broken symmetry generators equals the number of extra flat
directions in the Higgs potential. The Novikov theorem  is general,  it is
applicable irrespective of
whether the symmetry is true
or approximate (hidden), i.e.  irrespective of whether the NG
bosons are genuine or pseudo.

The Higgs potential, which is  'flat' along some directions (i.e. there are
rays in field space along which the potential vanishes) has been discussed
in relation to cosmological inflation and in supersymmetric theories, see
e.g. review \cite{FlatDirection}.
 As distinct from the other theories of flat directions, in $^3$He the
quartic terms in the Higgs potential in Eq. (\ref{HiggsPotential})  are
non-zero. Nevertheless, for some sub-manifold of vacuum states,
 the extra flat direction leads  to  9  NG modes for
8 broken symmetry generators.  The flat directions are 'lifted'
when the hidden symmetry is  violated, as a result  the quasi NG modes
acquire mass and become the Higgs bosons.

\section{Flat directions  in Higgs potential and extended $SO(6)$ symmetry}
\label{FlatDirections}

 Here we demonstrate, how the extra flat directions lead to substantial
extension of the symmetry of tangent space.
For that we add  two components of spin singlet $s$-wave Higgs field $\Psi$
to 18 components of spin-triplet $p$-wave Higgs field $A_{\alpha i}$, and
introduce the set of 15+1 generators of transformation or 15+1 operators:
\begin{equation}
{\cal I}~,~ {\cal L}_i~,~{\cal S}^{\alpha}~,~{\cal P}^{\alpha}_i
\label{16operators}
\end{equation}
The set contains the generators of the conventional group
$G=SO(3)_L\otimes
SO(3)_S\otimes U(1)$: three components of the  orbital angular
momentum, $ {\cal L}_i$ ($i=1,2,3$) (generators of the group
$SO(3)_L$ of orbital rotation) +  3 components of spin  angular momentum
${\cal
S}^\alpha$ ($\alpha=1,2,3$) (generators of the group
$SO(3)_S$ of spin rotation) + generator ${\cal I}$ of global $U(1)$ group of
phase rotations acting  on the Higgs fields as:
\begin{equation}
 {\cal L}_i A^\alpha_j =ie_{ijk}A^\alpha_k~~,~~{\cal S}^\alpha  A^\beta_i
=ie^{\alpha\beta\gamma}A^\gamma_i ~~,~~
 {\cal I}  A^\beta_i = A^\gamma_i~~,~~{\cal I} \Psi = \Psi \,.
\label{SA-commutators}
\end{equation}
The extended group has 9 more generators  ${\cal
P}^\alpha_i$, which act on Higgs fields as
\begin{equation}
 {\cal P}^\alpha_i  A^\beta_k = e^{\alpha\beta\gamma}e_{ijk}A^\gamma_k +
\Psi  \delta^{\alpha\beta}\delta_{ik} ~,~{\cal P}^\alpha_i \Psi
= A^\alpha_i \,.
\label{PA-commutators}
\end{equation}
New elements of symmetry mix triplet and singlet amplitudes of
Higgs field.

The nonzero commutators of these 16 operators are
\begin{equation}
[{\cal L}_i,{\cal L}_j]=ie_{ijk}{\cal L}_k ~~,~~
[{\cal S}^\alpha, {\cal S}^\beta]=ie^{\alpha\beta\gamma}{\cal S}^\gamma
\label{S-commutators}
\end{equation}
\begin{equation}
 [{\cal S}^\alpha, {\cal P}^\beta_i]=ie^{\alpha\beta\gamma}{\cal
P}^\gamma_i~~,~~
[{\cal L}_i,{\cal P}^\alpha_j]=ie_{ijk}{\cal P}^\alpha_k
\label{LP-commutators}
\end{equation}
\begin{equation}
[{\cal P}^\alpha_i, {\cal
P}^\beta_j]=i\left(\delta^{\alpha\beta}e_{ijk}{\cal L}_k +
\delta_{ij}e^{\alpha\beta\gamma}{\cal S}^\gamma \right)
\label{PP-commutators}
\end{equation}

The 15 generators in Eq.(\ref{16operators}) form the $SO(6)$
group:
\begin{equation}
[{\cal \lambda}_{ab},{\cal \lambda}_{bc}]=i{\cal \lambda}_{ca}
\label{lambda-commutators}
\end{equation}
where ${\cal \lambda}_{ab}$ is antisymmetric $6\times 6$ matrix with
components:
\begin{equation}
{\cal \lambda}_{12}={\cal L}_{z}~,~ {\cal \lambda}_{23}={\cal L}_{x}~,~{\cal
\lambda}_{31}={\cal L}_{y} ~~,~~
{\cal \lambda}_{45}={\cal S}^{z}~,~ {\cal \lambda}_{56}={\cal S}^{x}~,~{\cal
\lambda}_{64}={\cal S}^{y}~,
\label{lambda-generators2}
\end{equation}
\begin{equation}
{\cal \lambda}_{14}={\cal P}^x_x ~,~ {\cal \lambda}_{15}={\cal
P}^y_x ~,~  {\cal \lambda}_{16}={\cal P}^z_x
\label{lambda-generators3}
\end{equation}
\begin{equation}
{\cal \lambda}_{24}={\cal P}^x_y  ~,~ {\cal \lambda}_{25}={\cal P}^y_y  ~,~
{\cal
\lambda}_{26}={\cal P}^z_y
\label{lambda-generators4}
\end{equation}
\begin{equation}
 {\cal \lambda}_{34}={\cal P}^x_z ~,~ {\cal\lambda}_{35}={\cal P}^y_z ~,~
{\cal
\lambda}_{36}={\cal P}^z_z
\label{lambda-generators5}
\end{equation}
Together with  the gauge group $U(1)$, the hidden symmetry group in BCS
regime is $G_{\rm h}=SO(6)\otimes U(1)$, which transforms the Higgs field as
\begin{equation}
 (A^\alpha_i, \Psi)\rightarrow e^{i\phi {\cal I}}e^{i\theta^\alpha {\cal
S}^\alpha } e^{i\theta_i {\cal
L}_i } e^{i\Omega^\beta_k {\cal
P}^\beta_k }(A^\alpha_i,
\Psi)
\label{transformation}
\end{equation}
Here $\theta^\alpha$ and $\theta_i$ are rotation angles in spin and orbital
spaces correspondingly, $\phi$ is the parameter of the phase rotations,
while 9
other parameters $\Omega^\beta_k$ are angles of additional rotations of
$SO(6)$ group. Thus 10 complex components of the triplet +singlet Higgs form
10 dimensional representation of the $SU(4)$ or $SO(6)$
group.

This extended symmetry group describes the properties of the $^3$He-A in the
BCS approximation, if
 $\hat{\bf l}_+=\hat{\bf l}_-$ and the vacuum state of the Higgs field is
$A^\alpha_i({\rm eq})=\Delta_0 \hat x^\alpha(\hat x_i +i\hat y_i)$.
The expansion of the Higgs potential  in terms of the deviations of the
Higgs field from its equilibrium value in Eq.(\ref{B-phase})  is  (in
dimensionless units):
\begin{equation}
\delta F = \sum_\alpha[(u^\alpha_1 -v^\alpha_2)^2+ (u^\alpha_2
+v^\alpha_1)^2 ] +
2[(u^1_1+v^1_2)^2+(u^2_2-v^2_1)^2+(u^3_2-v^3_1)^2]
\label{QuadraticForm}
\end{equation}
This quadratic form is exactly zero if the deviations of the
order parameter are obtained by the action of all elements of $G_{\rm h}$,
i.e. if
$\delta A^\alpha_i(G_{\rm h})=G_{\rm h}A^\alpha_i({\rm eq})-
A^\alpha_i({\rm eq})$. Thus $G_{\rm h}$ is the extended symmetry of the
Higgs potential
in tangent space, and this symmetry leads to the flat
directions. Its subgroup $H_{\rm h}$ -- the symmetry group of the vacuum
state ($H_{\rm h}A^\alpha_i({\rm eq})=0$) --
has 5 generators:
\begin{equation}
H_{\rm h}=SU(2)\otimes U(1)\otimes U(1) \,,
\label{Hidden-group}
\end{equation}
\begin{equation}
\left({{\cal S}^z - {\cal P}^z_z\over 2}~, ~{{\cal S}^x - {\cal P}^x_z\over
2}~,~{{\cal S}^y - {\cal P}^y_z\over 2}\right)~;~ {\cal S}^x +{\cal
P}^x_z~;~{\cal I}-{\cal L}_z=Q\,,
\label{GenerationsHidden-group}
\end{equation}
where $Q$ is again the analog of electric  charge in SM.  So, the BCS model
of $^3$He-A contains $16-5=11$
NG bosons (two of them correspond to oscillations of the scalar
condensate $\Psi$) and $20-11=9$ Higgs modes.

The conventional symmetry breaking pattern in $^3$He-A, $G=SO_S(3)\otimes
SO_L(3)\otimes U(1) \rightarrow H =  SO_S(2)\otimes U_Q(1)$, gives $7-2=5$
NG bosons. The flat
directions emerging in the BCS model lead to 6 additional NG bosons, or to
4 if
one neglects the oscillations of the scalar Higgs field $\Psi$.
When the explicit corrections to the weak coupling approximation are
introduced, or the quantum corrections are taken into account, these 4 modes
become Higgs bosons with  small masses.
 See Ref. \cite{Saunders1989} for experiments with massive Higgs modes in
$^3$He-A.

\section{Nambu sum rule for gapless fermions}
\label{GaplessFermions}

Similar to the 2D case in Eqs.(\ref{2DA1}) and
(\ref{2DA2}), in 3D $^3$He-A the modes with ``electric charge'' $Q=\pm 2$
and $Q=0$ obey
the Nambu sum rule, but in a modified form. In 3D $^3$He-A, the  gap in
the fermionic spectrum is anisotropic and vanishes in the direction of
$\hat{\bf l}$.
The nodes in spectrum demonstrate another possible scenario of the symmetry
breaking in SM, which leads
to splitting of the degenerate Fermi point instead of formation of the
fermionic mass  \cite{KlinkhamerVolovik2005}.
The lesson from $^3$He-A is that in such case, the term $M_f^2$
 in the Nambu sum rule (\ref{NambuRelation}) must be substituted by the
angle average of the square of  anisotropic gap \cite{VolovikZubkov2013}.
For $^3$He-A one obtains
  \begin{eqnarray}
E_1^{(Q=0)} =0~~,~~ E_2^{(Q=0)} = 2\bar{\Delta} ~~,~~~
E^{(Q=+2)}=E^{(Q=-2)}= \sqrt{2} \bar{\Delta}\,,
\\
\bar{\Delta}^2 \equiv  \left<\Delta^2(\theta)\right> =\frac{2}{3}\Delta_0^2
\,.
\label{ClappingModes}
\end{eqnarray}

\section{Broken time reversal symmetry: Higgs from NG boson}
\label{TimeReversalSymmetry}

As is well known in condensed matter community, the violation of time
reversal symmetry $T$ leads to splitting of
NG bosons with linear spectrum to the mode with quadratic spectrum and the
mode with the
gapped spectrum, the Higgs mode. NG bosons with quadratic dispersion
correspond to two broken
generators while those with linear dispersion correspond to
one broken generator, see also recent discussion in
\cite{WatanabeMurayama2012}.
In particular, this happens for spin waves
in ferromagnets, where $T$ is spontaneously broken, and for  Kelvin waves
propagating along a  vortex in superfluids, where the  circulating
flow around the vortex breaks the $T$-symmetry.

In ferromagnets, the symmetry breaking pattern is $SO_S(3) \rightarrow
SO_S(2)$. Typically this leads to  $3-1=2$ NG modes with
linear spectrum $\omega_{1,2}=ck$ (spin waves). The broken $T$ symmetry
transforms the two branches  into quadratic NG mode
and the Higgs mode. For small $k$ one has
\begin{equation}
 \omega_1 =\frac{k^2}{M}~~,~~\omega_2 =Mc^2 + \frac{k^2}{M} ~~,~~k\ll Mc\,.
  \label{Ferromagnet}
\end{equation}
Superfluid  $^3$He-A has orbital angular momentum and thus represents the
liquid orbital ferromagnet.
Splitting of the linear spectrum of orbital waves in $^3$He-A according to
Eq.(\ref{Ferromagnet}) can be found in Eqs.(6.52-54) in \cite{Exotic}.
Orbital waves in  $^3$He-A  are analogs of photons. However, in SM  such
splitting would be possible only if the
CPT and Lorentz symmetries are violated.

A vortex line  in superfluids breaks translational symmetry in two
transverse directions.  The similar linear topological defect without
violation of $T$-symmetry would have two NG modes propagating along the
line. The broken $T$ symmetry of the vortex combines two NG modes with
linear spectrum  into one NG mode with quadratic spectrum -- the Kelvin
wave --
according to Eq.(\ref{Ferromagnet}). Recent discussion of the NG modes on
vortices and strings see in: \cite{ShifmanYung2013,NittaShifman2013}.

\section{Nambu sum rules in the relativistic models of top quark
condensation}
\label{NJL}

We consider the NJL model of general type that involves all $6$ quarks and
all $6$ leptons (neutrino is supposed to be of Dirac type). Let us consider
the particular form of the four  - fermion action. It is obtained assuming
that the tensor of coupling constants standing in front of the
four-fermion
term is factorized and  that lepton number originates
from the fourth color in the spirit of Pati-Salam models. The action of
the NJL model has the form
\begin{eqnarray}
S & = & \int d^4x \Bigl(\bar{\chi}[ i \nabla \gamma ]\chi  +
\frac{8\pi^2}{\Lambda^2} (\bar{\chi}_{k, \alpha A,L} \chi^{l, \beta,
B}_R)(\bar{\chi}_{\bar{l},\bar{{\beta}}
\bar{B}, R} {\chi}^{\bar{k},\bar{\alpha} {A}}_{L}) W^k_{\bar{k}}
W_l^{\bar{l}} L_{\bar{\alpha}}^{\alpha}
R_{\beta}^{\bar{\beta}} I_B^{\bar{B}}\Bigr) \label{Stopcolor_}
\end{eqnarray}

Here $\chi_{k, \alpha A}^T = \{(u_k,d_k); (c_k,s_k); (t_k,b_k)\}$ for $k =
1,2,3$ is the set of quark doublets
with the generation index $\alpha$, while $\chi_{4, \alpha A}^T = \{(\nu_e,
e); (\nu_{\mu},\mu); (\nu_{\tau},\tau)\}$ is the set of  lepton doublets.
$\Lambda$ is the dimensional parameter.
Hermitian matrices  $L,R,I,W$  contain dimensionless coupling constants. The
form of action Eq. (\ref{Stopcolor_}) with $W = {\rm diag} \, (1+
\frac{1}{2}W_{e\mu\tau},1,1,1)$ is fixed by the requirement that there is
the $SU(3)\otimes SU(2) \otimes U(1)$ symmetry.
We imply that all eigenvalues of matrices $L,R,I$ are close to each
other.
We assume the existence of an approximate symmetry: at the zero order of a
perturbation theory the eigenvalues of $L,R,I$ are all equal to each other,
and $W_{e\mu\tau}=0$. For example, the action of the corresponding form
appears in the model with the gauge field of Lorentz group \cite{Z2013}. Any
small corrections to this equality gives the eigenvalues of $L,R,I$ that
only
slightly deviate from each other, and the value of $W_{e\mu\tau}$ that only
slightly deviates from $0$. (After suitable rescaling $\Lambda$ plays
the role of the cutoff,
while the eigenvalues of $L,R,I$ are all close to $1$.)

Bosonic spectrum of this model is formally given by the expressions for the
bosonic spectrum of the model suggested in \cite{Miransky} and calculated in
one - loop approximation in
\cite{VolovikZubkov2013}. It is implied that in vacuum the composite scalar
fields $h_q = \bar{q}q$ are condensed for all fermions
$q=u,d,c,s,t,b,e,\mu,\tau, \nu_e,\nu_{\mu}, \nu_{\tau}$.

 There are two excitations in each $q\bar{q}$ channel with masses
$M^{P}_{q\bar{q}}$ and $M^{S}_{q\bar{q}}$
 and four excitations (i.e. two doubly degenerated excitations) in each
$q_1\bar{q}_2$ channel.  (Pairings of leptons and quarks are also allowed
and give the colored scalar fields.) We denote the masses
$M^{\pm}_{q_1\bar{q}_2},
M^{\pm}_{q_2\bar{q}_1}$. It is worth mentioning that each of the scalar
quark - antiquark bosons carries two color indices. In the absence of the
$SU(3)$ gauge field each of these channels represents the degenerate nonet.
When the color interactions are turned on we are left with the singlet and
octet states. Traceless octet states as well as the color scalar excitations
of the quark - lepton channels cannot exist as distinct particles due to
color confinement.

Instead of the trivial Nambu sum rule of the simplest models of top - quark
condensation $M_H = 2 M_t$ we have the sum rule \cite{VolovikZubkov2013}:
\begin{eqnarray}
&& [M^{+}_{q_1\bar{q}_2}]^2  + [M^{-}_{q_1\bar{q}_2}]^2 +
[M^{+}_{q_2\bar{q}_1}]^2  + [M^{-}_{q_2\bar{q}_1}]^2\approx 4 [M_{q_1}^2 +
M_{q_2}^2],  \quad (q_1\ne q_2); \nonumber\\
&& [M^{P}_{q\bar{q}}]^2  + [M^{S}_{q\bar{q}}]^2 \approx 4
M_{q}^2\label{NSRR}
\end{eqnarray}
In the case when the t-quark contributes to the formation of the given
scalar excitation,
its mass dominates, and in each channel ($t\bar{t}, t\bar{c}$, ...) we come
to the relation $\sum M_{H, i}^2  \approx 4 M^2_t$,
where the sum is over scalar excitations in the given channel.

It is important, that although the corrections to the eigenvalues of
$L,R,I,W$ are small, this does not mean that the corrections to the masses
are small. Instead, the large difference between masses may appear in this
way.
The symmetry breaking pattern of the considered model is $U_{ud,
L}(2)\otimes
...\otimes U(2)_{e \nu_e, L} \otimes U(1)_u \otimes ... \otimes U(1)_e
\rightarrow U(1)_u\otimes ...\otimes U(1)_e$. Among the
mentioned Higgs bosons there are 24 Goldstone bosons that are exactly
massless (in the channels $t(1\pm \gamma^5)\bar b, t \gamma^5\bar{t},
c(1\pm\gamma^5)\bar{s}, c\gamma^5\bar{c}, u(1\pm \gamma^5)\bar{d},
u\gamma^5\bar{u},b\gamma^5\bar{b}, s\gamma^5\bar{s}, d\gamma^5\bar{d}$ and
in the similar lepton - lepton channels).
There are Higgs bosons with the masses of the order of the t-quark mass ($
t(1\pm \gamma^5)\bar b, t \bar{t},  t(1\pm\gamma^5)\bar{s},
t\gamma^5\bar{c}, t(1\pm \gamma^5)\bar{d}, t\gamma^5\bar{u}$, and similar
quark - lepton states). The other
Higgs bosons have masses much smaller than the t - quark mass.
That's why a lot of physics is to be added in order to make this model
realistic. Extra light Higgs bosons should be provided with the masses of
the order of $M_t$. In principle, this may be achieved if the new gauge
symmetries are added, that are spontaneously broken. Then the extra light
Higgs bosons may become massive via the Higgs mechanism.

In principle, all Higgs bosons $h$ in the channels $t \bar{t},b \bar{b},
\bar{\tau} \tau, \bar{\nu}_{\tau} \nu_{\tau} ..., d \bar{d},t(1\pm
\gamma^5)\bar b, ... $ are coupled to the fields of the Standard Model in a
similar way. However, already at the tree level the corresponding coupling
constants are different for different Higgs bosons.
 (The form of the Higgs boson decay Lagrangian is given in  \cite{status}.)
The cross - sections of the processes (that may be observed at the LHC) like
$p p \rightarrow h \rightarrow WW, ZZ, gg, \gamma \gamma$ for the $\bar{t} t
$ Higgs bosons are much larger than for the other Higgs Bosons and are close
to that of the Standard Model.
This means, in particular, that the scalar boson of the present model in the
$\bar{t}t$ channel with mass $\approx 350$ GeV is excluded by the LHC data.
Therefore, some additional physics is necessary that either suppresses the
corresponding cross - section or makes this state much heavier.
The decays of the other Higgs bosons to $ZZ, WW, \gamma \gamma, gg$ are
suppressed compared to that of $\bar{t}t$. Therefore, these scalar states
are not excluded by the LHC data.
In the processes like $p p \rightarrow h \rightarrow \bar{c}c, \bar{b}b,
\bar{\tau} \tau$ the scalar states $\bar{c}c$, $\bar{t}{t}$,
$\bar{\tau}\tau$ dominate at the tree level. At the present moment we do not
comment on the possible exclusion of these states by the LHC data.

\section{Conclusions}

 Experience with the Higgs and NG bosons in condensed matter allows us to
suspect, that
 the observed Higgs boson is not fundamental: it may come as a composite
object emerging in the fermionic vacuum. If so, there can be several species
of Higgs bosons with different quantum numbers and with hierarchy of  masses
related to the hierarchy of hidden symmetries. Some particular analogies
with condensed matter allows us even to predict the possible values of
masses of extra Higgs bosons using the Nambu sum rule. The hint
from superfluid $^3$He-B suggests the mass
 $\sim 325$ GeV, while the hint from superfluid $^3$He-A suggests two
degenerated Higgs bosons with mass  $\sim 245$  GeV. However, in the
particular relativistic model of top quark condensation the four (two pairs)
Higgs bosons contribute to the sum rule of Eq. (\ref{NSRR}). This pattern
suggests the mass  $210$ GeV for the Nambu partner of the $125$ GeV Higgs
boson.

 In relation to cosmology, the thermodynamics of quantum liquids
allows us to explain why the huge vacuum energy of Higgs fields
does not contribute to cosmological constant in  equilibrium
\cite{KlinkhamerVolovik2008}.

It is worth mentioning that the Nambu relation between the masses of Higgs
bosons and the fermion masses is valid only in the one - loop approximation.
Formally, this approximation works in the relativistic NJL model only, when
the higher loop quadratic divergences are subtracted. At the present moment
the source of such a subtraction remains unclear. However,
there exists the theory, where in the similar situation it does takes place.
In quantum hydrodynamics \cite{quantum_hydro} there formally exist the
divergent contributions to various quantities  (say, to vacuum energy) due
to zero point energy of quantized sound waves -- phonons.
The quantum hydrodynamics is to be considered
as a theory with finite cutoff $\Lambda$. The loop divergences in the vacuum
energy are to be subtracted just like we do for the case of the NJL model.
In hydrodynamics the explanation of such a subtraction is that the
microscopic theory to which the hydrodynamics is an approximation works both
at the energies smaller and larger than $\Lambda$, and this microscopic
theory contains the contributions from the energies larger than $\Lambda$.
Due to the
thermodynamical stability of vacuum, these contributions exactly cancel the
divergences appeared in the low
energy effective theory. In \cite{hydro_gravity} it was
suggested that a similar pattern may provide the mechanism for the
cancellation of the divergent contributions to vacuum energy in quantum
gravity and divergent contributions to the Higgs boson mass in the Standard
Model. { We suppose, that in our case of the NJL model the contributions of
the trans - $\Lambda$ degrees of freedom cancel the dominant divergences in
the bosonic and fermionic masses leaving us with the one - loop
approximation as an effective tool for the evaluation of physical
quantities.}

\begin{acknowledgements}

This work was partly supported by RFBR grant 11-02-01227, by the
Federal Special-Purpose Programme 'Human Capital' of the Russian Ministry of
Science and Education. GEV
acknowledges a financial support of the Academy of Finland and its COE
program,
and the EU  FP7 program ($\#$228464 Microkelvin).

\end{acknowledgements}


\end{document}